\documentclass[preprint,aps,prl,epsfig]{revtex4}
\usepackage{graphicx} 
\usepackage{dcolumn}
\usepackage{bm}

\begin{document}

\title{Analytical model for tracer dispersion in porous media}
\author{B.Ph. van Milligen$^1$ and P.D. Bons$^2$}
\address{$^1$CIEMAT, Avda. Complutense 40, 28040 Madrid, Spain\\
$^2$Mineralogie und Geodynamik, Eberhard Karls Universit\"at, Wilhelmstrasse 56, 72074 T\"ubingen, Germany}
\date{\today}
\pacs{81.05.Rm,66.10.cg,47.56.+r}

\begin{abstract}
In this work, we present a novel analytical model for tracer dispersion in laminar flow through porous media.
Based on a straightforward physical argument, it describes the generic behavior of dispersion
over a wide range of P\'eclet numbers (exceeding 8 orders of magnitude).
In particular, the model accurately captures the intermediate scaling behavior of longitudinal dispersion, obviating the need to subdivide the dispersional behavior into a number of disjunct regimes or using empirical power law expressions.

The analysis also reveals the existence of a new material property, the critical P\'eclet number, which reflects the mesoscale geometric properties of the microscopic pore {structure.}

\end{abstract}
\maketitle
\section{Introduction}
Solute dispersion in porous media is of importance in many fields of science, such as chemistry, groundwater hydrology, oil recovery, etc. Tracers, dissolved in a fluid flowing through a porous material, will experience dispersion in both the longitudinal (downstream) and transverse directions, due to thermal or Fickian diffusion and the variability in both flow velocity and direction in the pores. Tracers injected at one point (e.g., a leaking tank polluting groundwater) will thus spread out into a plume. 

Over the past decades, many models have been proposed to model tracer transport. 
Most models relate the dispersion coefficients to the flow velocity ($v$), a characteristic length scale (the grain size or pore length $G$), and the molecular diffusion coefficient ($D_m$) via  the dimensionless P\'eclet number Pe $= vG/D_m$. 
Due to the use of the microscale $G$,  the P\'eclet number is a microscopic property. 
The most basic approach to computing the spatiotemporal evolution of the concentration of tracers is provided by the advection-diffusion equation (ADE), using separate values of $D_L$ and $D_T$ for the dispersion coefficients in the longitudinal and transverse direction, respectively.
To explain the apparently anomalous behavior of the dispersion coefficients observed in experiments, more sophisticated models have been developed, e.g., flow through random capillaries~\cite{Saffman:1959}.
The analysis of~\cite{Bear:1967} provided an initial approach to evaluate the macroscopic effects of microscopic pore geometry by means of an averaging procedure.
Percolation theory and Continuous Time Random Walks allowed handling long range correlations~\cite{Sahimi:1993}. 
A range of numerical methods have also been applied to the problem, such as network models in which pore connectivity plays an essential role (e.g., \cite{Bijeljic:2007}). 
Finally, full microscopic modeling of incompressible flow through porous materials has become feasible thanks to modern computing, and considerable successes have been obtained with this method (e.g., \cite{Maier:2000,Mostaghimi:2010,Ovaysi:2011}).

In this paper we propose a new approach to understand the first-order relationship between dispersion and flow velocity (and hence P\'eclet number). {It is based on} a straightforward physical argument involving the competition of diffusion and advection in the pore channels, and leads to an accurate prediction of the observed behavior over the full relevant range of P\'eclet numbers. The analysis also reveals the existence of a new material property related to the cited competition and the mesoscopic structure of pore geometry.

\section{Modeling}

\subsection{Longitudinal dispersion}

Consider the laminar flow of fluid through a regular or random homogeneous porous material.
The flow is produced by a pressure difference (head) applied at opposite ends of the sample of porous material. This pressure difference induces a complex flow pattern inside the pores that can be computed exactly using the appropriate equations for an incompressible fluid, with appropriate boundary conditions, cf.~\cite{Ovaysi:2011}.

Due to the complexity of the pore structure, the flow lines go apart and come together again as they traverse the material, 
dictated by the pressure head and the pore structure, so that the 
dispersion of the longitudinal distance  $\left < (\Delta d_L)^2\right >$ among different flow lines tends to grow linearly as a function of the mean longitudinal distance $\left < d_L\right >$, taken in the direction of the negative pressure gradient.
Regardless of the complexity of the flow pattern inside the porous material, the incompressible flow at every point is linear in the applied pressure head, so the geometry of this flow pattern does not vary as the head is varied.
This leads to what is known as `mechanical' dispersion, with a diffusion coefficient given by
\begin{equation}\label{mechanical}
D_L = \frac{\left < (\Delta d_L)^2\right >}{\Delta t} = \frac{\left < (\Delta d_L)^2\right >}{\left < d_L\right >/\left <v_L\right >} =  \beta_L \left < v_L\right >,
\end{equation}
where the mean longitudinal flow velocity $\left < v_L \right >= v$ has been taken as a measure for the pressure head, 
and $\beta_L$ is a geometric proportionality constant.
Thus, the longitudinal dispersion $D_L$ is linear in the mean longitudinal flow velocity $v$.
A similar argument can also be applied to the transverse dispersion, that is likewise linear in the flow velocity, 
but with a different (smaller) proportionality constant due to the significant geometrical angle between the direction of the mean pressure gradient and the direction of the transport: $D_T = \left <(\Delta d_T)^2\right >/\Delta t =  \beta_T \left < v_L\right >$.

Into this flow, tracers are injected. These tracers are not merely advected with the flow,  but in addition are subject to random (thermal) Brownian motion.
At zero or very small flow, pure thermal diffusion will result, with diffusion coefficient $D=D_0$. 
At large flow velocity, but still in the laminar flow regime, the thermal motion of the tracers can be neglected with respect to the advection, and nearly pure `mechanical' dispersion will result, $D=D_L$ or $D_T$.

We expect there to be an intermediate or transitional regime, in which global tracer dispersion should lie somewhere between the thermal and mechanical diffusion limits, but {\it a priori} it is not clear how exactly this dispersion scales with flow velocity.

Following a tracer path {through the pore channels of the porous material}, there will be channels oriented at a small angle to the driving force, in which the mechanical dispersion will tend to dominate over thermal diffusion. Other channels will be oriented nearly at right angles to the driving force, and there, thermal diffusion will tend to dominate over mechanical diffusion.
Thus, we propose modeling the mixed behavior in this intermediate regime by assuming that transport in each of the successive pore channels traversed by a tracer is dominated by one of the two mechanisms.
In other words, we assume that a  tracer alternately experiences thermal and mechanical diffusion.
Thus, the tracer experiences thermal diffusion for a fraction of time $t_0$, with thermal diffusion coefficient $D_0$.  
Alternating with these phases of thermal diffusion, the tracer is advected by the flow for a fraction of time $t_L$, during which it experiences a mechanical dispersion $D_L$ (proportional to $v$).
Therefore, the total dispersion of the longitudinal distance is the weighted sum of these two:
\begin{eqnarray}
\left < (\Delta d_L)^2\right > = D_0 t_0 + D_L t_L, \nonumber
\end{eqnarray}
so the net or total longitudinal diffusion is
\begin{eqnarray}
D^t_L = \frac{\left <(\Delta d_L)^2\right >}{t_0+t_L} = \frac{D_0t_0+ D_L t_L }{t_0 + t_L} = \frac{D_0 + D_L (t_L/t_0)}{1+t_L/t_0}. \nonumber
\end{eqnarray}
Now, as the net longitudinal flow velocity $v$ increases, we make the essential assumption that the time fraction $t_L$ becomes progressively longer with respect to $t_0$ as $v$ increases.
This is motivated as follows.
{As the drive is increased, transport in ever more pore channels becomes dominated by mechanical dispersion.}
So the time fraction ratio of mechanical to thermal dispersion a tracer experiences 
{will be an increasing function of the velocity:
\begin{eqnarray}
\frac{t_L}{t_0} = f\left ( \frac{v}{v_c} \right ), \nonumber
\end{eqnarray}
where $f(x)$ is a monotonically increasing function of $x$ such that $f(0)=0$ and $f(1)=1$, and}
$v_c$ is a critical velocity (the velocity where mechanical diffusion starts to dominate globally over thermal diffusion).
In the limit $v=0$, one has $t_L/t_0=0$, and $D^t_L=D_0$. 
In the limit $v\to \infty$, one has $t_L/t_0 \to \infty$, and $D^t_L=D_L$. 
Both these limits agree with expectation.
{In the following, we will set $f(x)=x$ for simplicity, but leave open the possibility that} 
{future studies may reveal more complex functional dependencies.}
To better understand the intermediate regime, we insert $D_L = \beta_L v$, and find
\begin{equation}\label{longitudinal}
D^t_L = \frac{D_0 + \beta_L v f(v/v_c)}{1+f(v/v_c)}
= \frac{D_0 + \beta_L v (v/v_c)}{1+v/v_c}
\end{equation}
{which is recognized as a Pad\'e approximation.}

Eq.~(\ref{longitudinal}) has {\it triple} asymptotic behavior.
For $v \downarrow 0$, one has $D^t_L \simeq D_0$.
For $v \to \infty$, one has $D^t_L \simeq \beta_L v$.
But there is also an intermediate regime where $D^t_L \simeq \beta_L v^2/v_c$.
This regime occurs for $v/v_c \ll 1$ and $\beta_L v^2/v_c \gg D_0$. Summarizing:
\begin{eqnarray}
D^t_L \simeq \left\{\begin{array}{ll}
D_0 & \mathrm{when~} \frac{v}{v_c} \ll \sqrt{\frac{D_0}{\beta_Lv_c}} \\
\beta_L v^2/v_c & \mathrm{when~} \sqrt{\frac{D_0}{\beta_Lv_c}}\ll \frac{v}{v_c} \ll 1 \\
\beta_L v & \mathrm{when~} \frac{v}{v_c} \gg 1 \end{array}\right. \nonumber
\end{eqnarray}
The intermediate asymptote only appears in full when the two corresponding limits  are sufficiently far apart, i.e., when
$D_0/\beta_Lv_c \ll 1$. 

Physically, the left and right asymptotes correspond to the limits in which one of the transport mechanisms (thermal diffusion and mechanical dispersion, respectively) dominates.
The intermediate regime arises when the strength of the mechanical dispersion increases simultaneously with the fraction of time $t_L/t_0$ that the tracer is experiencing mechanical dispersion as compared to thermal diffusion. 

Experimental data are commonly expressed in terms of $D^t/D_m$ versus Pe, where $D_m$ is the molecular diffusion coefficient and Pe is the dimensionless P\'eclet number. The above expressions can be recast in this dimensionless form by substituting $D^t \to D^t/D_m$ and $\{v,v_c\} \to$ \{Pe,Pe$_c$\}. {Thus, the dimensionless form of Eq.~(\ref{longitudinal}) is:
\begin{equation}\label{longitudinal_DL}
\frac{D^t_L}{D_m} = \frac{D_0/D_m + \beta'_L \mathrm{Pe} ( \mathrm{Pe}/ \mathrm{Pe}_c)}{1+ \mathrm{Pe}/ \mathrm{Pe}_c}
\end{equation}
}

\begin{figure}
  \includegraphics[trim=0 0 0 0,clip=,width=16cm]{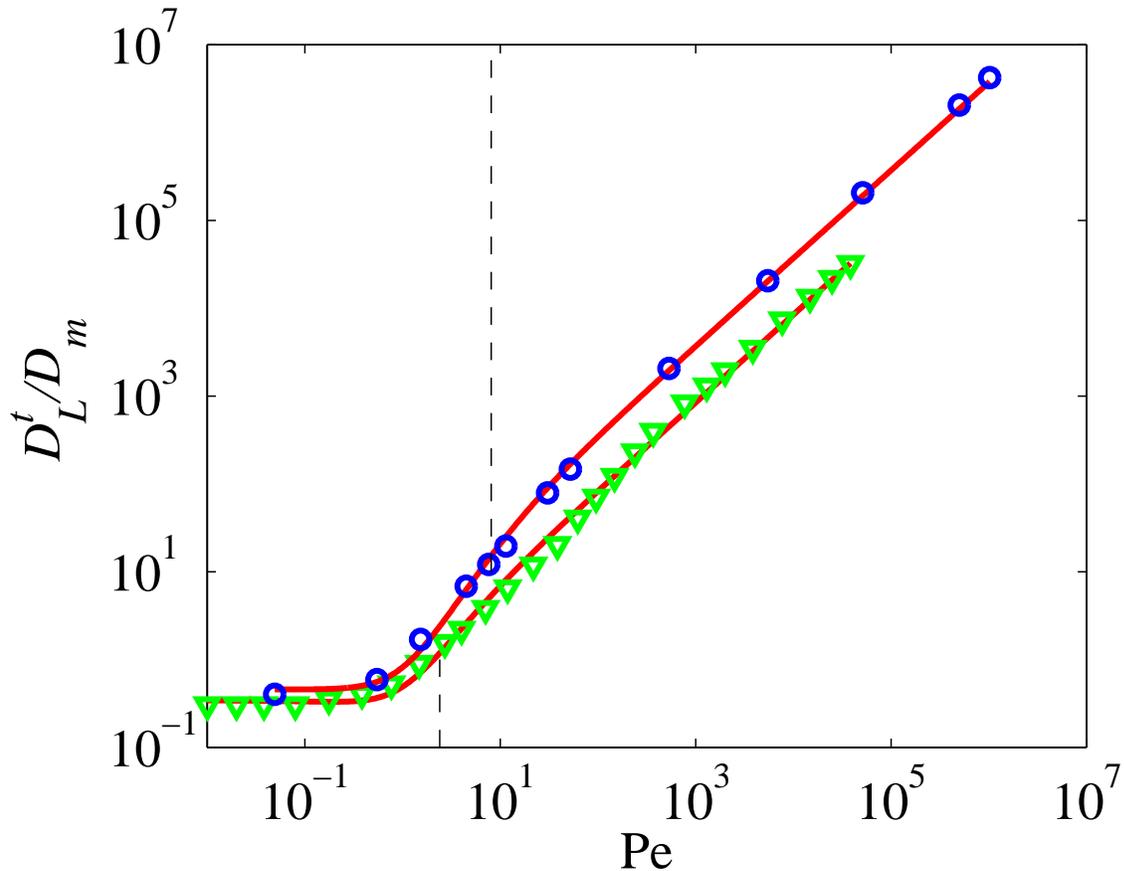}
\caption{\label{Ovaysi}{Numerical simulations of longitudinal dispersion in 3D micro-CT scans of sandstone samples, and model curves.} {Top curve:} Data from Sample A of \cite{Ovaysi:2011} (blue circles); line: fit using {Eq.~(\ref{longitudinal_DL}) with $D_0/D_m = 0.46 \pm 0.13$, Pe$_c=8.0 \pm 0.2$, and $\beta'_L=3.7 \pm 0.5$}. 
Bottom curve: Data from \cite{Mostaghimi:2010} (green triangles); line: fit using {Eq.~(\ref{longitudinal_DL}) with $D_0/D_m = 0.34 \pm 0.05$, Pe$_c=2.4 \pm 0.3$, and $\beta'_L=0.84 \pm 0.09$.
The vertical dashed lines indicate Pe$_c$ for each curve.}}
\end{figure}

To test the validity of Eq.~(\ref{longitudinal_DL}), we have fitted our expression to the numerical simulation data of {\cite{Mostaghimi:2010} and \cite{Ovaysi:2011}  (based on micro computer tomography (CT) scans of porous sandstone)}.
Due to the very low noise level of the numerical simulation, this test is {probably} more stringent than testing the model against data from laboratory experiments.
The {fits}, shown in Fig.~\ref{Ovaysi}, {are} a very good match to the simulated data. Note, in particular, that the intermediate region (with logarithmic slope $> 1$) is reproduced in full {detail.}

\subsection{Transverse dispersion}

Transverse dispersion is different from longitudinal dispersion in that the drive is not very efficient in the transverse direction. 
Therefore, the transverse mechanical dispersion is not able to compete with thermal diffusion, and thermal diffusion cannot be neglected along any part of the tracer trajectory. 
The drive, however, is only effective in pore channels directed at angles less than 90$^\circ$ with respect to the mean flow vector, during a fraction of time $t_L$.
Thus, we get
\begin{eqnarray}
\left <(\Delta d_T)^2 \right > = D_0(t_0+t_L) + D_T t_L, \nonumber
\end{eqnarray}
which expresses that thermal diffusion is always operative in the transverse direction, while mechanical dispersion (with characteristic diffusion coefficient $D_T$) is operative only in specific channels, i.e., for a fraction of the total time.
Due to the relative inefficiency of the drive, this approach is expected to yield less precise results than Eq.~(\ref{longitudinal}), and minor corrections to this expression may be needed~\cite{Klenk:2002}. This is left to future work.

Following the same reasoning as before, we immediately find
\begin{equation}\label{transverse}
D^t_T = \frac{\left <(\Delta d_T)^2\right >}{t_0+t_L} = D_0 + D_T \frac{t_L/t_0}{1+t_L/t_0} = 
D_0 + \beta_Tv \frac{v/v_c}{1+ v/v_c} 
\end{equation}

We have assumed that $t_L$ (or $v_c$) has the same value here as with longitudinal dispersion, based on the assumption that the geometrical flow pattern is one and the same for longitudinal and transverse dispersion, although this is only strictly true in the limit of small thermal diffusion.

The transverse dispersion has two extreme asymptotes that are equivalent to those of longitudinal dispersion, 
namely: for $v \downarrow 0$, one has $D^t_T \simeq D_0$, and
for $v \to \infty$, one has $D^t_T \simeq \beta_T v$.
However, here there is no `intermediate asymptote', but only a very gradual transition from one asymptote to the other, specified via the factor containing $v/v_c$ in Eq.~(\ref{transverse}).
The transition between asymptotes occurs around $v \simeq v_c$.

{
In dimensionless form, Eq.~(\ref{transverse}) becomes:
\begin{equation}\label{transverse_DL}
\frac{D^t_T}{D_m} =\frac{D_0}{D_m} + \beta'_T\mathrm{Pe} \frac{\mathrm{Pe}/\mathrm{Pe}_c}{1+ \mathrm{Pe}/\mathrm{Pe}_c} 
\end{equation} }

\begin{figure}
  \includegraphics[trim=0 0 0 0,clip=,width=16cm]{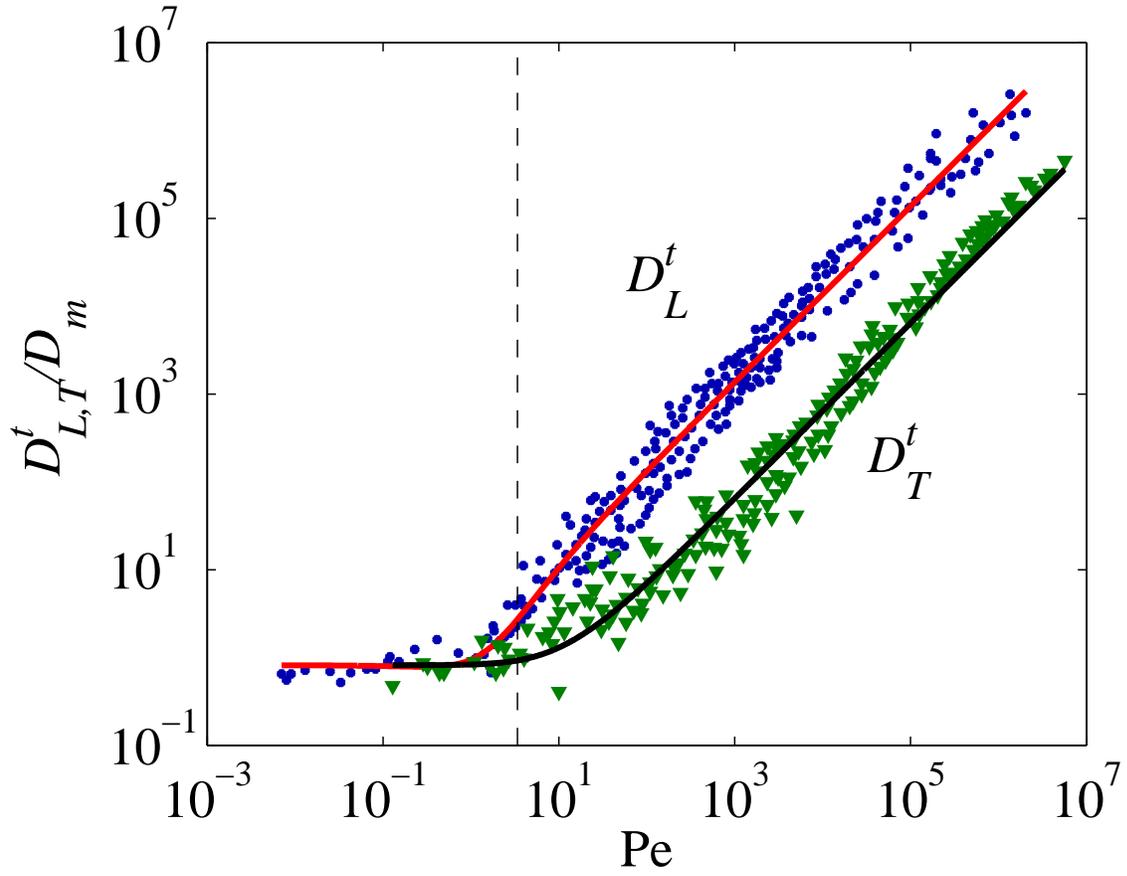}
\caption{\label{Delgado}Longitudinal (circles) and transverse (triangles) dispersion. Data {from experiments in a wide range of granular porous media} from \cite{Delgado:2007} and references therein.
The longitudinal dispersion data set was fitted using {Eq.~(\ref{longitudinal_DL}), yielding $D_0/D_m = 0.8 \pm 0.2$, Pe$_c=3.4\pm 0.2$, and $\beta'_L=1.4\pm0.1$}. The transverse dispersion data set was fitted using {Eq.~(\ref{transverse_DL}) while keeping $D_0/D_m$ and Pe$_c$ fixed, yielding $\beta'_T=0.064 \pm 0.006$}. The vertical dashed line indicates Pe$_c$.}
\end{figure}

A fit of the analytic expressions to an ample collection of measurement data is shown in Fig.~\ref{Delgado}.
{The figure contains data from 18 different experiments on longitudinal dispersion and 15 on transverse dispersion, spanning an ample range of the Schmidt number, $500 < Sc < 2000$~\cite{Delgado:2007}.}
The parameters {$D_0/D_m$, Pe$_c$ and $\beta'_L$} were determined from a least-squares fit to the longitudinal dataset.
For the transverse dataset, the parameters {$D_0/D_m$} and Pe$_c$ were held fixed at these values, while only {$\beta'_T$} was varied.
{By making a joint fit to all data, the obtained parameters represent mean values over the various individual datasets. This procedure illustrates the generic dispersional behavior, covering a range of different materials.
But much better results, with less scatter, are to be expected for fits to an individual dataset.}

\begin{figure}
  \includegraphics[trim=0 0 0 0,clip=,width=16cm]{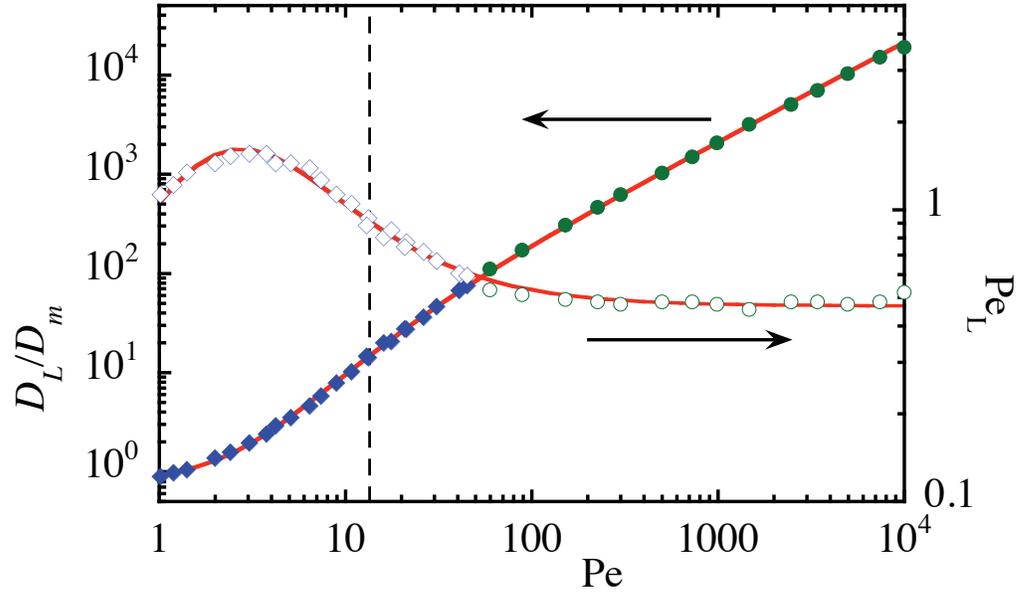}
\caption{\label{Carvalho}{Longitudinal dispersion in a packed bed of glass beads. Diamonds: data from \cite{Carvalho:2005} at $Sc=754$. Circles: data from \cite{Carvalho:2003} at $Sc=424$.
Closed symbols: $D_L/D_m$. Open symbols: the same data represented as $\mathrm{Pe}_L \equiv \mathrm{Pe}D_m/D_L$.
The continuous curves correspond to the fit of the data to Eq.~(\ref{longitudinal_DL}), yielding $D_0/D_m = 0.86 \pm 0.05$, Pe$_c=13.5\pm 1$, and $\beta'_L=2.13\pm0.05$. The vertical dashed line indicates Pe$_c$. }}
\end{figure}

{To illustrate the preceding remark, Fig.~\ref{Carvalho} shows an example of a fit to an individual longitudinal dispersion dataset. The fit is rather satisfactory.}

\clearpage

\section{Discussion}

In previous work \cite{Sahimi:1993,Delgado:2007}, the available experimental data were analyzed by subdividing the range of Pe numbers into individual regimes and describing the dispersional behavior in each of these sections heuristically (using, e.g., power-law expressions~\cite{Olsson:2007}).

\begin{figure}
  \includegraphics[trim=0 0 0 0,clip=,width=12cm]{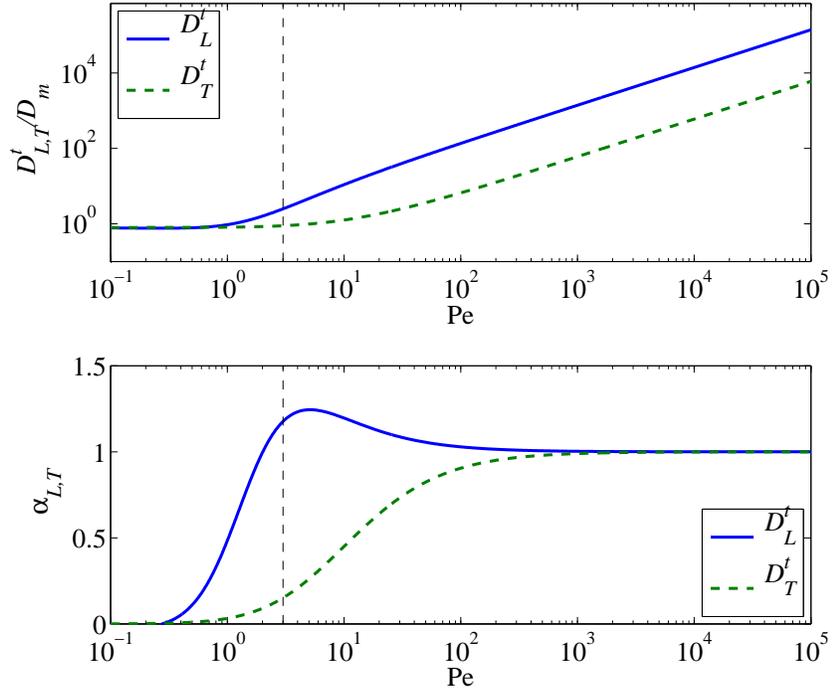}
\caption{\label{Exponent}The two analytical curves for longitudinal and transverse dispersion, $D^t_L(\mathrm{Pe})$ and $D^t_T(\mathrm{Pe})$, and their local power-law exponent given by Eq.~(\ref{power-law}).}
\end{figure}

In Fig.~\ref{Exponent} we show (top) the typical shape of the analytical curves, Eqs.~(\ref{longitudinal_DL}) and (\ref{transverse_DL}), with parameters similar to those used in the examples above ({$D_0/D_m=0.8$, Pe$_c= 3$, $\beta'_L=1.4$, $\beta'_T=0.06$}).
The bottom panel in this figure shows the local power-law exponent, computed as 
\begin{equation}\label{power-law}
\alpha_{L,T} = \frac{\mathrm{Pe}}{D^t_{L,T}}\frac{\partial D^t_{L,T}}{\partial \mathrm{Pe}}
\end{equation}
The evolution of the local power-law exponent with Pe number displays roughly the same behavior as the successive regimes described in \cite{Sahimi:1993,Delgado:2007}.
The intermediate regime does not develop fully as the condition $(D_0/D_m)/\beta'_L$Pe$_c \ll 1$ is only marginally fulfilled.
Thus, the intermediate power-law exponent for longitudinal dispersion only reaches a value of about 1.2, instead of the maximum of 2 predicted by the model.
This value (1.2) is consistent with the value cited in literature for the corresponding range of Pe numbers~\cite{Sorbie:1991,Delgado:2007}.

\clearpage
\section{Conclusions}

In this work, we present a unified model for longitudinal and transverse tracer dispersion in laminar flow in porous media.
Based on a straightforward physical argument, an analytical expression is obtained that describes the observed behavior over the full available range of laminar flow velocities or P\'eclet numbers.

In literature, it has been customary to use power-law scalings to describe the behavior of $D^t_{L,T}(\mathrm{Pe})$, each of which being valid only in a limited range of Pe values~\cite{Sahimi:1993,Delgado:2007}. The generic appearance of power-law scalings seemed to indicate that tracer dispersion in porous media was an anomalous transport process, characterized by fractional exponents~\cite{Sahimi:1993}. However, in view of the present work, it seems that no anomalous transport mechanisms need be invoked to explain the observed behavior. This is a very satisfactory situation, at least for sandbox or glass bead experiments, in which the structure of the medium is not patently fractal.

The present work has revealed the existence of two numbers that determine the dispersion curves: the critical velocity or P\'eclet number ($v_c$ or Pe$_c$) and $D_0/\beta_L\mathrm{Pe}_c$. 
Pe$_c$ is a material property that depends on the pore geometry, so that different materials will have different Pe$_c$ values. 
{The exploration of the dependence of Pe$_c$ on other material properties may provide further insight into its significance.}
It is of interest to note that we generally find that Pe$_c>1$, rather than equal to one, as one might expect for the transition from dominant diffusive to dominant advective transport~\cite{Bijeljic:2007}. This indicates that the scale at which this transition occurs is larger (by a factor Pe$_c$) than the grain or pore scale $G$.
{Also note that the family of three-parameter curves given by $D_0, \mathrm{Pe}_c$ and $\beta_L$ may explain a significant part of the observed data spread mentioned in e.g.~\cite{Delgado:2007}.}
Finally, the intermediate region and the maximum value of the local longitudinal power-law exponent depend exclusively on $D_0/\beta_L\mathrm{Pe}_c$.

The analysis we present is consistent with the idea, proposed also by other authors~\cite{Hunt:2010}, that the main effects of porosity on dispersion can be modelled using a single characteristic mesoscopic {length} scale (Pe$_cG$), and that the details of the microscopic pore geometry and inertial effects only lead to corrections to this generic behavior.
{The present modeling approach allows determining this mesoscale from the dispersion data.}
The present work does not pretend to provide the same modeling precision as the empirical scalings provided in the ample literature, but merely to provide improved insight into the {generic} behavior of tracer dispersion in driven laminar flow through porous media.

\section*{Acknowledgements}

This study was carried out within the framework of DGMK (German Society for Petroleum and Coal Science and Technology) research project 718 ``Mineral Vein Dynamics Modelling'', which is funded by the companies ExxonMobil Production Deutschland GmbH, GDF SUEZ E\&P Deutschland GmbH, RWE Dea AG and Wintershall Holding GmbH, within the basic research program of the WEG Wirtschaftsverband Erd\"ol- und Erdgasgewinnung e.V. We thank the companies for their financial support and their permission to publish these results. 
This research was partially supported by grant ENE2009-07247, Ministerio de Ciencia e Innovaci\'on (Spain).

\bibliographystyle{apsrev}

\begin{thebibliography}{14}
\expandafter\ifx\csname natexlab\endcsname\relax\def\natexlab#1{#1}\fi
\expandafter\ifx\csname bibnamefont\endcsname\relax
  \def\bibnamefont#1{#1}\fi
\expandafter\ifx\csname bibfnamefont\endcsname\relax
  \def\bibfnamefont#1{#1}\fi
\expandafter\ifx\csname citenamefont\endcsname\relax
  \def\citenamefont#1{#1}\fi
\expandafter\ifx\csname url\endcsname\relax
  \def\url#1{\texttt{#1}}\fi
\expandafter\ifx\csname urlprefix\endcsname\relax\def\urlprefix{URL }\fi
\providecommand{\bibinfo}[2]{#2}
\providecommand{\eprint}[2][]{\url{#2}}

\bibitem[{\citenamefont{Saffman}(1959)}]{Saffman:1959}
\bibinfo{author}{\bibfnamefont{P.}~\bibnamefont{Saffman}}, \bibinfo{journal}{J.
  Fluid Mech.} \textbf{\bibinfo{volume}{6}}, \bibinfo{pages}{321}
  (\bibinfo{year}{1959}).

\bibitem[{\citenamefont{Bear and Bachmat}(1967)}]{Bear:1967}
\bibinfo{author}{\bibfnamefont{J.}~\bibnamefont{Bear}} \bibnamefont{and}
  \bibinfo{author}{\bibfnamefont{Y.}~\bibnamefont{Bachmat}},
  \bibinfo{journal}{Proc. I.A.S.H. Symposium on Artificial Recharge and
  Management of Aquafers} \textbf{\bibinfo{volume}{IASH Publ. 72}},
  \bibinfo{pages}{7} (\bibinfo{year}{1967}).

\bibitem[{\citenamefont{Sahimi}(1993)}]{Sahimi:1993}
\bibinfo{author}{\bibfnamefont{M.}~\bibnamefont{Sahimi}},
  \bibinfo{journal}{Transport in Porous Media} \textbf{\bibinfo{volume}{13}},
  \bibinfo{pages}{3} (\bibinfo{year}{1993}).

\bibitem[{\citenamefont{Bijeljic and Blunt}(2007)}]{Bijeljic:2007}
\bibinfo{author}{\bibfnamefont{B.}~\bibnamefont{Bijeljic}} \bibnamefont{and}
  \bibinfo{author}{\bibfnamefont{M.}~\bibnamefont{Blunt}},
  \bibinfo{journal}{Water Resources Research} \textbf{\bibinfo{volume}{43}},
  \bibinfo{pages}{W12S11} (\bibinfo{year}{2007}).

\bibitem[{\citenamefont{Maier et~al.}(2000)\citenamefont{Maier, Kroll, Bernard,
  Howington, Peters, and Ted~Davis}}]{Maier:2000}
\bibinfo{author}{\bibfnamefont{R.}~\bibnamefont{Maier}},
  \bibinfo{author}{\bibfnamefont{D.}~\bibnamefont{Kroll}},
  \bibinfo{author}{\bibfnamefont{R.}~\bibnamefont{Bernard}},
  \bibinfo{author}{\bibfnamefont{S.}~\bibnamefont{Howington}},
  \bibinfo{author}{\bibfnamefont{J.}~\bibnamefont{Peters}}, \bibnamefont{and}
  \bibinfo{author}{\bibfnamefont{H.}~\bibnamefont{Ted~Davis}},
  \bibinfo{journal}{Phys. Fluids} \textbf{\bibinfo{volume}{12}},
  \bibinfo{pages}{2065} (\bibinfo{year}{2000}).

\bibitem[{\citenamefont{Mostaghimi et~al.}(2010)\citenamefont{Mostaghimi,
  Bijeljic, and Blunt}}]{Mostaghimi:2010}
\bibinfo{author}{\bibfnamefont{P.}~\bibnamefont{Mostaghimi}},
  \bibinfo{author}{\bibfnamefont{B.}~\bibnamefont{Bijeljic}}, \bibnamefont{and}
  \bibinfo{author}{\bibfnamefont{M.}~\bibnamefont{Blunt}},
  \bibinfo{journal}{Proc. SPE Annual Technical Conference and Exhibition,
  Florence, Italy} \textbf{\bibinfo{volume}{SPE}}, \bibinfo{pages}{135261}
  (\bibinfo{year}{2010}).

\bibitem[{\citenamefont{Ovaysi and Piri}(2011)}]{Ovaysi:2011}
\bibinfo{author}{\bibfnamefont{S.}~\bibnamefont{Ovaysi}} \bibnamefont{and}
  \bibinfo{author}{\bibfnamefont{M.}~\bibnamefont{Piri}}, \bibinfo{journal}{J.
  Contaminant Hydrology} \textbf{\bibinfo{volume}{124}}, \bibinfo{pages}{68}
  (\bibinfo{year}{2011}).

\bibitem[{\citenamefont{Klenk and Grathwohl}(2002)}]{Klenk:2002}
\bibinfo{author}{\bibfnamefont{I.}~\bibnamefont{Klenk}} \bibnamefont{and}
  \bibinfo{author}{\bibfnamefont{P.}~\bibnamefont{Grathwohl}},
  \bibinfo{journal}{J. Contaminant Hydrology} \textbf{\bibinfo{volume}{58}},
  \bibinfo{pages}{111} (\bibinfo{year}{2002}).

\bibitem[{\citenamefont{Delgado}(2007)}]{Delgado:2007}
\bibinfo{author}{\bibfnamefont{J.}~\bibnamefont{Delgado}},
  \bibinfo{journal}{Trans. IChemE Part A} \textbf{\bibinfo{volume}{85}},
  \bibinfo{pages}{1245} (\bibinfo{year}{2007}).

\bibitem[{\citenamefont{Guedes~de Carvalho and Delgado}(2005)}]{Carvalho:2005}
\bibinfo{author}{\bibfnamefont{J.}~\bibnamefont{Guedes~de Carvalho}}
  \bibnamefont{and} \bibinfo{author}{\bibfnamefont{J.}~\bibnamefont{Delgado}},
  \bibinfo{journal}{Chem. Eng. Science} \textbf{\bibinfo{volume}{60}},
  \bibinfo{pages}{365} (\bibinfo{year}{2005}).

\bibitem[{\citenamefont{Guedes~de Carvalho and Delgado}(2003)}]{Carvalho:2003}
\bibinfo{author}{\bibfnamefont{J.}~\bibnamefont{Guedes~de Carvalho}}
  \bibnamefont{and} \bibinfo{author}{\bibfnamefont{J.}~\bibnamefont{Delgado}},
  \bibinfo{journal}{A.I.Ch.E. Journal} \textbf{\bibinfo{volume}{49}},
  \bibinfo{pages}{1980} (\bibinfo{year}{2003}).

\bibitem[{\citenamefont{Olsson and Grathwohl}(2007)}]{Olsson:2007}
\bibinfo{author}{\bibfnamefont{A.}~\bibnamefont{Olsson}} \bibnamefont{and}
  \bibinfo{author}{\bibfnamefont{P.}~\bibnamefont{Grathwohl}},
  \bibinfo{journal}{J. Contaminant Hydrology} \textbf{\bibinfo{volume}{92}},
  \bibinfo{pages}{149} (\bibinfo{year}{2007}).

\bibitem[{\citenamefont{Sorbie and Clifford}(1991)}]{Sorbie:1991}
\bibinfo{author}{\bibfnamefont{K.}~\bibnamefont{Sorbie}} \bibnamefont{and}
  \bibinfo{author}{\bibfnamefont{P.}~\bibnamefont{Clifford}},
  \bibinfo{journal}{Chem. Eng. Science} \textbf{\bibinfo{volume}{46}},
  \bibinfo{pages}{2525} (\bibinfo{year}{1991}).

\bibitem[{\citenamefont{Hunt and Skinner}(2010)}]{Hunt:2010}
\bibinfo{author}{\bibfnamefont{A.}~\bibnamefont{Hunt}} \bibnamefont{and}
  \bibinfo{author}{\bibfnamefont{T.}~\bibnamefont{Skinner}},
  \bibinfo{journal}{Complexity} \textbf{\bibinfo{volume}{16}},
  \bibinfo{pages}{43} (\bibinfo{year}{2010}).

\end{thebibliography}

\end{document}